\input{aipcheck}

\documentclass[
    ,draft           
  ]
  {aipproc}

\layoutstyle{6x9}

\def\be{\begin{equation}}
\def\ee{\end{equation}}
\def\ba{\begin{eqnarray}}
\def\ea{\end{eqnarray}}

\begin{document}

\title{Noncanonical Phase-Space Noncommutative Black Holes}

                

\author{Catarina Bastos}{
  address={Instituto de Plasmas e Fus\~ao Nuclear, Instituto Superior T\'ecnico,\
  Avenida Rovisco Pais 1, 1049-001 Lisboa, Portugal},
  ,email={ catarina.bastos@ist.utl.pt}
}

\author{Orfeu Bertolami}{ 
 address={Departamento de F\'isica e Astronomia,\
 Faculdade de Ci\^encias da Universidade do Porto,\
 Rua do Campo Alegre, 687, 4169-007 Porto, Portugal}
,altaddress={Instituto de Plasmas e Fus\~ao Nuclear, Instituto Superior T\'ecnico,\
  Avenida Rovisco Pais 1, 1049-001 Lisboa, Portugal},
  ,email={orfeu.bertolami@fc.up.pt}
}

\author{Nuno Costa Dias}{
  address={Departamento de Matem\'atica ,\
  Universidade Lus\'ofona de Humanidades e Tecnologias,\
  Avenida Campo Grande, 376, 1749-024 Lisboa, Portugal}
  ,altaddress={Grupo de F\'isica Matem\'atica, UL, Avenida Prof. Gama Pinto 2, 1649-003, Lisboa, Portugal.},
  ,email={ncdias@meo.pt}
   }
  
  \author{Jo\~ao Nuno Prata}{
  address={Departamento de Matem\'atica ,\
  Universidade Lus\'ofona de Humanidades e Tecnologias,\
  Avenida Campo Grande, 376, 1749-024 Lisboa, Portugal}
  ,altaddress={Grupo de F\'isica Matem\'atica, UL, Avenida Prof. Gama Pinto 2, 1649-003, Lisboa, Portugal.},
  ,email={joao.prata@mail.telepac.pt}
   }

\begin{abstract}
  In this contribution we present a noncanonical phase-space noncommutative (NC) extension of a Kantowski Sachs (KS) cosmological model to describe the interior of a Schwarzschild black hole (BH). We evaluate the thermodynamical quantities inside this NC Schwarzschild BH and compare with the well known quantities. We find that for a NCBH the temperature and entropy have the same mass dependence as the Hawking quantities for a Schwarzschild BH.
  \end{abstract}

\maketitle


\section{Introduction}

We consider a noncanonical phase-space NC extension of a Kantowski Sachs (KS) cosmological model to sudy the interior of a Schwarzschild BH \cite{Bastos6,Bastos8}. In a Schwarzschild BH, for $r<2M$, time and radial coordinates are interchanged ($r\leftrightarrow t$) and space-time is described by the metric:
\be\label{eq0.2}
ds^2=-\left({2M\over t}-1\right)^{-1}dt^2+\left({2M\over t}-1\right)dr^2+t^2(d\theta^2+\sin^2\theta d\varphi^2)~.
\ee
Hence, it can be seen that the interior of this BH can be described by an anisotropic cosmological space-time, as it is the KS cosmological model \cite{AK}, which in the Misner parametrization is given by
\be\label{eq1.4}
ds^2=-N^2dt^2+e^{2\sqrt{3}\beta}dr^2+e^{-2\sqrt{3}(\beta+\Omega)}(d\theta^2+\sin^2{\theta}d\varphi^2)~,
\ee
where $\Omega$ and $\beta$ are scale factors, and $N$ is the lapse function. The following identification for $t<2M$,
\be\label{eq0.3}
N^2=\left({2M\over t}-1\right)^{-1} \quad, \quad
e^{2\sqrt{3}\beta} = \left({2M\over t}-1\right) \quad , \quad
e^{-2\sqrt{3}\beta}e^{-2\sqrt{3}\Omega}=t^2~,
\ee
allows for mapping the metric Eq. (\ref{eq1.4}) into the metric Eq. (\ref{eq0.2}).

Let us consider that the scale factors $\Omega$ and $\beta$ and the corresponding conjugated momenta $P_{\Omega}$ and $P_{\beta}$ do not commute,
\ba\label{eq1.5}
\left[\hat{\Omega}, \hat{\beta} \right]\! &=&\! i \theta \left( 1 + \epsilon \theta \hat{\Omega} + {\epsilon \theta^2\over{1 + \sqrt{1- \xi}}} \hat{P}_{\beta} \right)\\
\left[\hat{P}_{\Omega}, \hat{P}_{\beta} \right]\! &=&\! i \left( \eta  + \epsilon (1 + \sqrt{1 - \xi})^2 \hat{\Omega} + \epsilon \theta (1 + \sqrt{1- \xi}) \hat{P}_{\beta} \right)\nonumber\\
\left[\hat{\Omega}, \hat{P}_{\Omega} \right]\! &=&\! \left[\hat{\beta}, \hat{P}_{\beta} \right]\! =\!i  \left( 1 + \epsilon \theta (1 + \sqrt{1-
\xi}) \hat{\Omega} + \epsilon \theta^2 \hat{P}_{\beta} \right)\nonumber,
\ea
where $\theta$, $\eta$ and $\epsilon$ are positive constants and $\xi = \theta \eta <1$ \cite{Bastos7}. The remaining relations vanish. In Ref. \cite{Bastos6}, the noncommutative Wheeler-deWitt (NCWDW) equation of a KS cosmological model was obtained and is used here to study the interior of a Schwarzschild BH. When $\epsilon=0$ we recover the canonical NC relations as in Refs. \cite{Bastos1,Bastos2,Bastos5}. To relate this NC variables with the usual commutative ones, we employ the following transformations \cite{Bastos6}
\ba\label{eq1.7}
\hat{\Omega} \!\!&=&\!\! \lambda \hat{\Omega}_c -  {\theta\over2\lambda} \hat{P}_{\beta_c} + E \hat{\Omega}_c^2\hspace{0.2cm},\hspace{0.2cm}\hat{\beta} = \lambda \hat{\beta}_c + {\theta\over2 \lambda} \hat{P}_{\Omega_c} \nonumber\\
\hat{P}_{\Omega}\!\! &=&\!\! \mu \hat{P}_{\Omega_c} +  {\eta\over2\mu} \hat{\beta}_c \hspace{0.2cm},\hspace{0.2cm}\hat{P}_{\beta} = \mu \hat{P}_{\beta_c} -  {\eta\over2 \mu} \hat{\Omega}_c + F \hat{\Omega}_c^2.
\ea
Here, $\mu$, $ \lambda$ are real parameters such that $(\lambda \mu)^2 - \lambda \mu + \frac{\xi}{4} =0\Leftrightarrow 2 \lambda \mu = 1 \pm \sqrt{1- \xi}$, and one chooses the positive solution given the invariance of the physics on the above transformations \cite{Bastos1}, and
\be\label{eq1.9}
E = - {\theta\over{1 + \sqrt{1- \xi}}} F\;,\;F = - {\lambda\over\mu} \epsilon \sqrt{1- \xi} \left(1 + \sqrt{1- \xi} \right)~.
\ee
Of course, the inverse transformation can be obtained as discussed in Ref. \cite{Bastos6}.

The NCWDW equation is obtained considering the canonical quantization of the KS Hamiltonian, and the fact that the operator $\hat A =\mu  \hat{P}_{\beta_c} + {\eta\over2 \mu} \hat{\Omega}_c$ commutes with the Hamiltonian (it corresponds to a constant of motion of the classical problem), and writing the wave function as $ \psi_a\left( \Omega_c, \beta_c \right)= {\cal R} \left( \Omega_c \right)\exp \left[ {i \beta_c\over\mu} \left(a - {\eta\over2 \mu}\Omega_c \right) \right]$ \cite{Bastos6,Bastos8}
\ba\label{eq1.19}
&&\mu^2 {\cal R}''- 48 \exp \left( - {2 \sqrt{3}\over\mu} \Omega_c - 2 \sqrt{3} E \Omega_c^2 + {\sqrt{3} \theta a\over\mu \lambda} \right) {\cal R}- {2 \eta\over\mu} \left(\Omega_c+F\Omega_c^3\right){\cal R} + \nonumber\\&&+F^2 \Omega_c^4{\cal R} + a^2{\cal R} + \left({\eta^2\over\mu^2}+2 a F \right) \Omega_c^2{\cal R}=0~,
\ea
where we consider the transformations, Eqs. (\ref{eq1.7}). The dependence on $\beta_c$ has completely disappeared and we are left with an ordinary differential equation for
${\cal R} \left(\Omega_c \right)$. Through the substitution, $\Omega_c = \mu z$, ${d^2\over d \Omega_c^2} = {1\over\mu^2} {d^2\over d z^2}$ and ${\cal R} \left( \Omega_c (z) \right) := \phi_a (z)$
we obtain a second order linear differential equation, actually a Schr\"odinger-like equation:
\be\label{eq1.21}
- \phi_a'' (z) +V(z) \phi_a (z)= 0~,
\ee
where the potential function, V(z), reads:
\ba\label{eq1.22}
V(z) = \!\!-\!\! \left( \eta z -a \right)^2\!\!\!\! - F^2 \mu^4 z^4\!\!\! - 2  F \mu^2(\eta z-a) z^2\!\!\! +48 \exp \left(\!\! -2 \sqrt{3} z\!\! -\!\!2 \sqrt{3} \mu^2 E z^2 +\! {\sqrt{3} \theta a \over\mu \lambda} \right).
\ea
Equation (\ref{eq1.21}) depends explicitly on the noncommutative parameters $\theta$, $\eta$, $\epsilon$ and the eigenvalue $a$, $\hat{A}\psi=a\psi$. Notice that $E>0$ and that the potential, Eq. (\ref{eq1.22}), has a minimum only if $\eta\neq0$.

\section{Thermodynamics of NCBH}

In order to use the Feynman-Hibbs procedure to obtain the quantum correction to the potential Eq. (\ref{eq1.22}) and to get the relevant thermodynamical quantities, we must have a minimum for the potential function, which is possible only when the momenta do not commute. Moreover, the potential function has its minimum for low values of $z$ \cite{Bastos8} and thus, can be written as:
\be\label{eq2.1}
V(z)=-(\eta z-a)^2 + 48\exp{\left[-2\sqrt{3}(z- { \theta a \over 2 \mu \lambda}+\mu^2 Ez^2)\right]}~.
\ee

Following the procedure outlined in Ref. \cite{Bastos5}, we expand the exponential term in the potential Eq. (\ref{eq2.1}) to second order in powers of the $z-z_0$ variable. The minimum $z_0$ is defined by the derivative of the potential function \cite{Bastos8}. Introducing the expanded potential function into Eq. (\ref{eq1.21}) and comparing with the Schr\"odinger equation of an harmonic oscillator we find that the NC potential is given by
\be\label{eq2.9}
V_{NC}(y)=24(B-l^2)\left(y^2+{\beta_{BH}\over12}\right)~.
\ee
in terms of the variable $y\equiv z-z_0$, where $l:=\eta/4\sqrt{3}$ and $B:=6k[(1+2\mu^2Ez_0)^2-(\sqrt{3}/3)\mu^2E]-l^2$. 

Thus, we obtain the NC partition function, the NC internal energy of the BH and the NC BH temperature,
\be\label{eq2.13}
T_{BH}={4 \over M}(B-l^2)~,
\ee
where we have assumed that $M>>1$ and dropped terms proportinal to $M^{-2}$. This quantity is positive and for $E=0$, we recover the results obatined in Ref. \cite{Bastos5}. Thus, new corrections to the NC temperature are obtained when we impose the noncanonical algebra Eqs. (\ref{eq1.5}). Furthermore, it is important to refer that we still have the same mass dependence of the NC temperature as in the Hawking temperature for a Schwarzschild BH, $T_{BH}= {1/( 8 \pi M)}$. In fact, the Hawking temperature is recovered 
\be\label{eq2.13a}
z_0=2.26369 \hspace{1 cm} \eta_0=0.487~,
\ee
once setting $\theta=0.1$, $\epsilon=0.3$ and $a=18.89$. The value $\eta_0=0.487$, can be seen as a reference which allows us to get the Hawking temperature, and as $\eta$ increases, we have a gradual NC deformation of the Hawking temperature. 

Finally, the NC entropy of the BH is, neglecting as before terms proportional to $M^{-2}$ as $M>>1$,
\ba\label{eq2.15}
S_{BH}&\simeq&{M^2\over8\left(6k[(1+2\mu^2Ez_0)^2-{\sqrt{3}\over3}\mu^2E]-l^2\right)}\nonumber\\
&-&{1\over2}\ln\left({3M^2\over 6k[(1+2\mu^2Ez_0)^2-{\sqrt{3}\over3}\mu^2E]-l^2}\right)~.
\ea
Again, for the reference value $\eta=\eta_0$, we recover the Hawking entropy, plus some ``stringy" logarithmic correction.

\section{Conclusions}
In this contribution we have obtained the temperature and entropy of a Schwarzschild BH in a noncanonical NC phase-space. This is a  generalization of the commutative, and the canonical phase-space NC models. We have shown that the Hawking temperture and entropy can be recovered for a non-vanishing value of $\eta_0$, namely $\eta_0=0.487$.

It is worth pointing out that one of the main features of this model is the regularization of the Schwarzschild singularity \cite{Bastos6}, a feature that can be extended for the KS singularity as well. The regularization of the two types of singularities is due to the eigenstates of the assymptotic behaviour of the obtained wave functions, which are square integrable, even when they are associated to a continuous set of eigenstates \cite{Bastos6}. Actually, this is a property shared by all Hamiltonians with a potential that behaves like $V(z)\approx-Kz^{2+\epsilon}$ for some $K,\epsilon>0$ as $z\rightarrow\infty$ \cite{Gitman}.


\begin{theacknowledgments}
  The work of CB is supported by Funda\c{c}\~{a}o para a Ci\^{e}ncia e a Tecnologia (FCT) under the grant SFRH/BPD/62861/2009. The work of OB is partially supported by the FCT grant PTDC/FIS/111362/2009. NCD and JNP were partially supported by the grants PTDC/MAT/69635/2006 and PTDC/MAT/099880/2008 of FCT.
\end{theacknowledgments}



\bibliographystyle{aipprocl} 

\end{document}